\documentclass[11pt,a4paper]{article}
\usepackage{jheppub_kim}
\usepackage{rotating}
\usepackage{graphicx,epsfig}
\usepackage{amsmath}
\usepackage {amssymb}
\usepackage{subfigure}

\usepackage{amsfonts}
\usepackage{bm}

\usepackage{relsize}

\begin{document}

\title{Inflationary Dynamics with a Smooth Slow-Roll to Constant-Roll Era Transition}
\author[a,b]{S.D. Odintsov}
\author[c,d]{V.K. Oikonomou}

\affiliation[a]{ ICREA, Passeig Luis Companys, 23, 08010 Barcelona,
Spain}

\affiliation[b] {Institute of Space Sciences (IEEC-CSIC) C. Can
Magrans s/n, 08193 Barcelona, Spain}

\affiliation[c]{Laboratory for Theoretical Cosmology, Tomsk State
University of Control Systems and Radioelectronics (TUSUR), 634050
Tomsk, Russia}

\affiliation[d]{Tomsk State Pedagogical University, 634061 Tomsk,
Russia}

\abstract{ In this paper we investigate the implications of having a
varying second slow-roll index on the canonical scalar field
inflationary dynamics. We shall be interested in cases that the
second slow-roll can take small values and correspondingly large
values, for limiting cases of the function that quantifies the
variation of the second slow-roll index. As we demonstrate, this can
naturally introduce a smooth transition between slow-roll and
constant-roll eras. We discuss the theoretical implications of the
mechanism we introduce and we use various illustrative examples in
order to better understand the new features that the varying second
slow-roll index introduces. In the examples we will present, the
second slow-roll index has exponential dependence on the scalar
field, and in one of these cases, the slow-roll era corresponds to a
type of $\alpha$-attractor inflation. Finally, we briefly discuss
how the combination of slow-roll and constant-roll may lead to
non-Gaussianities in the primordial perturbations. }

%PACS numbers: 04.50.Kd, 95.36.+x, 98.80.-k, 98.80.Cq
%\pacs{04.50.Kd, 95.36.+x, 98.80.-k, 98.80.Cq,11.25.-w}

\maketitle

%\makeatletter
%\renewcommand{\theequation}{\Roman{section}\,\Alph{subsection}.\arabic{equation}}
%\@addtoreset{equation}{subsection}{section}
%\makeatother

%\makeatletter
%\renewcommand{\theequation}{\Roman{section}.\arabic{equation}}
%\@addtoreset{equation}{section}
%\makeatother

\def\pp{{\, \mid \hskip -1.5mm =}}
\def\cL{\mathcal{L}}
\def\be{\begin{equation}}
\def\ee{\end{equation}}
\def\bea{\begin{eqnarray}}
\def\eea{\end{eqnarray}}
\def\tr{\mathrm{tr}\, }
\def\nn{\nonumber \\}
\def\e{\mathrm{e}}

\section{Introduction}

Inflation is considered one of the most successful descriptions of
the early Universe
\cite{inflation1,inflation2,inflation3,inflation4,inflation5}, since
it offered consistent remedies to the old Big Bang cosmological
description. The terminology inflation actually characterizes a
period of acceleration in the early Universe, and this early-time
acceleration era can be realized by various theoretical frameworks
\cite{Nojiri:2006ri,Capozziello:2011et,Capozziello:2010zz,Nojiri:2010wj,Clifton:2011jh},
with the most frequently used being the canonical scalar field
description \cite{inflation1,inflation2}. According to the latest
observational data coming from the Planck collaboration
\cite{planck}, the so-called inflaton, which is the canonical scalar
field, must follow a slow-roll trajectory in a plateau like
potential for large values of the inflaton field. Actually, a recent
research stream indicated that these large plateau potentials belong
to a class of potentials which all predict the same spectral index
of primordial curvature perturbations and the same scalar-to-tensor
ratio. These are known as $\alpha$-attractor models
\cite{alpha1,alpha2,alpha3,alpha4,alpha5,alpha6,alpha7,alpha8,alpha9,alpha10,alpha10a,alpha11,linderefs1,linder,vernov,lindelast,attractorsI,attractorsII,dimop}
and quite well known and also viable models, such as the Starobinsky
model \cite{starob1,starob2} and the Higgs model of inflation
\cite{higgs}, actually belong to this class of potentials. Hence,
the canonical scalar field realization of inflation has quite
appealing attributes which mainly are to our opinion, the
compatibility with the observations and also the holistic and
appealingly simple description of inflation that the
$\alpha$-attractor models offer.

However one drawback of the single field inflation models is that
these models do not leave any ``free space'' for future
observational data that predict non-Gaussianities in the primordial
density perturbations. Actually, the Gaussian property of the
primordial modes is an assumption stemming from the fact that the
modes are considered uncorrelated (for a review on non-Gaussianities
see \cite{Chen:2010xka}). Up to date, observations show that the
primordial density perturbations have a Gaussian distribution, but
the cosmic variance is a considerable limitation for any existing
detection method. In the future, if the observations of the Cosmic
Microwave Background (CMB) anisotropy and of the galaxy distribution
are combined, then it is possible to detect non-Gaussianities in the
perturbations. In effect, this will call into question the canonical
scalar field and also slow-rolling models of inflation. Also, in
conjunction with future observations of non-Gaussianities, the fact
that low multipoles are suppressed in the CMB spectrum \cite{planck}
makes compelling to modify the single scalar field models in order
to make them robust against future observations.

One way to allow non-Gaussianities to occur in the predicted
spectrum by single scalar field models, is to abandon the slow-roll
condition, and a recent important research stream adopted this
approach
\cite{Inoue:2001zt,Tsamis:2003px,Kinney:2005vj,Tzirakis:2007bf,Namjoo:2012aa,Martin:2012pe,Motohashi:2014ppa,Cai:2016ngx,Motohashi:2017aob,Hirano:2016gmv,Anguelova:2015dgt,Cook:2015hma,Kumar:2015mfa}.
These models are called constant-roll models
\cite{Motohashi:2014ppa}, or fast-roll models \cite{Martin:2012pe},
but for the purpose of simplicity in this paper we shall refer to
these as ``constant-roll'' models. These scenarios are quite
appealing, since in some cases, viability with observations can be
achieved \cite{Motohashi:2017aob}, but also non-Gaussianities are
predicted in these theories \cite{Namjoo:2012aa,Martin:2012pe}. In
addition, for some alternative works on reconstruction of
inflationary potentials from the slow-roll indices, see
\cite{asxet1,asxet2}.

Due to the phenomenological importance of constant-roll models, but
also due to the successes of slow-roll models, in this paper we
shall investigate how it is possible to smoothly combine these two
eras, in the context of single canonical scalar field theory.
Particularly, we shall demonstrate how it is possible to describe a
smooth dynamical transition between these two eras. This will be
achieved by allowing the second slow-roll index $\eta$ to be a
general function of the scalar field, appropriately chosen so that
both eras are realized. Obviously the condition $\eta \ll 1$
realizes the slow-roll era, and when $\eta$ becomes of the order
$\eta \sim \mathcal{O}(1)$, then the constant-roll era commences. We
shall present various models, with the constant-roll era occurring
before or after the slow-roll era. The most appealing case generates
a slow-roll era which corresponds to the $\alpha$-attractor
potentials
\cite{alpha1,alpha2,alpha3,alpha4,alpha5,alpha6,alpha7,alpha8,alpha9,alpha10,alpha10a,alpha11,linderefs1,linder,vernov}
at leading order. We also support numerically our analytical
approach, and as we demonstrate, at least in the context of our
models, the constant-roll inflation must occur after the slow-roll
era, since in the opposite case, the resulting evolution does not
have a unique attractor, and the initial conditions of the scalar
field crucially alter the final attractor. This feature however is
very model dependent as we shall demonstrate.

Throughout this work we shall assume that the geometric background
is a flat Friedmann-Robertson-Walker metric of the form, \be
\label{metricfrw} ds^2 = - dt^2 + a(t)^2 \sum_{i=1,2,3}
\left(dx^i\right)^2\, , \ee where $a(t)$ is the scale factor.
Moreover, we assume that the connection is a symmetric, metric
compatible and torsion-less affine connection, the Levi-Civita
connection.

This paper is organized as follows: In section II we shall present
the formalism and the basic features of the dynamical transition
mechanism, and we discuss various modifications of single scalar
field inflationary dynamics, that this mechanism brings along. In
section III we present a first model for which the transition
between slow-roll and constant-roll eras is achieved, and we discuss
the various features of it. Also we investigate if the resulting
solution is an attractor of the cosmological system, both
numerically and analytically. Also the slow-roll era is analyzed in
some detail, and as we show, the resulting observational indices are
identical to the ones corresponding to $\alpha$-attractor models. In
section IV we discuss another similar scenario which realizes the
constant-roll to slow-roll transition, and the stability of the
solution is also investigated in detail.

\section{Dynamical Transition Between Constant-Roll Inflation and Slow-Roll Inflation: Formalism and Basic Features}

Consider a canonical scalar field in a flat FRW Universe, with
action,
 \be
\label{canonicalscalarfieldaction}
\mathcal{S}=\sqrt{-g}\left(\frac{R}{2}-\frac{1}{2}\partial_{\mu}\phi
\partial^{\mu}\phi -V(\varphi))\right)\, ,
\ee with $V(\varphi)$ being the canonical scalar field potential.
The energy density is equal to, \be
\label{energydensitysinglescalar}
\rho=\frac{1}{2}\dot{\varphi}^2+V(\varphi)\, , \ee and also the
pressure is,
 \be \label{pressuresinglescalar}
P=\frac{1}{2}\dot{\varphi}^2-V(\varphi)\, . \ee Hence, the resulting
Friedmann equation is, \be \label{friedmaneqnsinglescalar}
H^2=\frac{1}{3 M_p^2}\rho\, , \ee and we easily find that,
 \be
\label{dothsinglescalar} \dot{H}=-\frac{1}{2M_p^2}\dot{\varphi}^2\,
. \ee Also the canonical scalar field satisfies the Klein-Gordon
equation,
 \be \label{kleingordonsingle}
\ddot{\varphi}+3H\dot{\varphi}+V'=0\, , \ee with the prime denoting
differentiation with respect to the canonical scalar field
$\varphi$.

The inflationary dynamics are quantified in terms of the slow-roll
parameters $\epsilon$ and $\eta$, which are the lowest order
perturbation parameters in the so-called Hubble slow-roll expansion
\cite{Liddle:1994dx}. These are defined as follows, \be
\label{slowrollindiceshubblerate} \epsilon=-\frac{\dot{H}}{H^2}\,
,\quad \eta=-\frac{\ddot{H}}{2H\dot{H}}\, , \ee in terms of the
Hubble rate, and can be written in terms of the canonical scalar
field as follows, \be \label{slowrollindiceshubblerate123}
\epsilon=\frac{\dot{\varphi}^2}{2M_p^2H^2}\, ,\quad
\eta=-\frac{\ddot{\varphi}}{2H\dot{\varphi}}\, . \ee The
constant-roll inflation models
\cite{Martin:2012pe,Motohashi:2014ppa,Motohashi:2017aob} are based
on the fact that the second slow-roll index $\eta$, is not small
during the inflationary era, but it is constant, so it satisfies
$\eta=-n$, where $n$ is a constant parameter \cite{Martin:2012pe}.
Essentially, this condition eliminates the slow-roll era, which is
based on the assumption that during inflation, the slow-roll indices
satisfy $\epsilon,\eta \ll 1$, and the various phenomenological
implications of the condition $\eta=-n$, were studied in detail in
the literature. In this work we shall assume that the second
slow-roll index satisfies the following condition,
\begin{equation}\label{basciccnd1}
\eta=-f(\varphi (t))\, ,
\end{equation}
or equivalently,
\begin{equation}\label{basiccondition}
\frac{\ddot{\varphi}}{2H\dot{\varphi}}=f(\varphi (t))\, .
\end{equation}
The function $f(\varphi (t))$ in both Eqs. (\ref{basciccnd1}) and
(\ref{basiccondition}) is assumed to be a smooth and monotonic
function of the canonical scalar field $\varphi (t)$, which we shall
specify in the following sections. The condition
(\ref{basiccondition}) is the main assumption we shall make in this
paper, and we shall see that the transition from a slow-roll era to
a constant-roll era is controlled effectively by the condition
(\ref{basiccondition}). As we shall demonstrate, in all the cases
the transition is smooth.

Our strategy is to find the Hubble rate as a function of the
canonical scalar field $H(\varphi )$, by using the Hamilton-Jacobi
formalism, by solving the equations of motion. We need to note that,
as we discuss also later on, the solution $H(\varphi)$ must be
checked numerically and analytically in order to validate if it is
the attractor of the cosmological dynamical system, by using various
initial conditions. Coming back to the cosmological system, since
$\dot{H}=\dot{\varphi}\frac{\mathrm{d}H}{\mathrm{d}\varphi}$ we can
write Eq. (\ref{dothsinglescalar})  as follows,
\begin{equation}\label{extra1}
\dot{\varphi}=-2M_p^2\frac{\mathrm{d}H}{\mathrm{d}\varphi}\, ,
\end{equation}
so by differentiating Eq. (\ref{extra1}) with respect to the cosmic
time, and by substituting the result in Eq. (\ref{basiccondition}),
we obtain the following differential equation,
\begin{equation}\label{masterequation}
\frac{\mathrm{d}^2H}{\mathrm{d}\varphi^2}=-\frac{1}{2M_p^2}f(\varphi
)H(\varphi )\, .
\end{equation}
The above differential equation will be the master differential
equation in this paper, and it will determine the possible
attractors of the cosmological system. The scalar potential
$V(\varphi)$ can be expressed in terms of $H(\varphi)$, if we
substitute Eq. (\ref{extra1}) in Eq.
(\ref{friedmaneqnsinglescalar}), so we get,
\begin{equation}\label{potentialhubblscalar}
V(\varphi)=3M_p^2H(\varphi)^2-2M_p^4(H'(\varphi))^2\, .
\end{equation}
By solving the differential equation (\ref{masterequation}), we can
substitute the resulting solution $H(\varphi)$ in Eq.
(\ref{potentialhubblscalar}), and obtain the scalar potential
$V(\varphi)$.

However, a solution $H(\varphi)$ may not necessarily be an attractor
of the cosmological dynamics, so this should be checked both
analytically and numerically. For later convenience we discuss here
how to check analytically whether a solution $H_0(\varphi)$ is an
attractor of the cosmological system. Following
\cite{Liddle:1994dx},  we consider the variation of Eq.
(\ref{potentialhubblscalar}), and we obtain the following equation,
\begin{equation}\label{perturbationeqnsbbasic1}
H_0'(\varphi)\delta H'(\varphi)\simeq
\frac{3}{2M_p^2}H_0(\varphi)\delta H(\varphi)\, ,
\end{equation}
which can be solved given $H_0(\varphi)$, and it has the solution,
\begin{equation}\label{perturbationsolution1}
\delta H(\varphi )=\delta
H(\varphi_0)e^{\frac{3}{2M_p^2}\int_{\varphi_0}^{\varphi}\frac{H_0(\varphi)}{H_0(\varphi
)}}\, ,
\end{equation}
where $\varphi_0$ some initial value of the scalar field.
Effectively, for a solution $H_0(\varphi)$ of the differential
equation (\ref{masterequation}), we can check if the solution is an
attractor of the cosmological dynamics, if the linear perturbations
(\ref{perturbationsolution1}) decay or grow. However, a resulting
solution can be considered an attractor if also a numerical analysis
reveals an attractor behavior in the phase space
$(\dot{\varphi}(t),\varphi (t))$. In all the examples we shall
study, we shall use both the analytic and the numerical approach.

In order to have transitions between the slow-roll and constant-roll
era, it is conceivable that the function $f(\varphi (t))$ must be
chosen in such a way so that during the slow-roll era it satisfies
$f(\varphi (t)) \simeq 0$, and during the constant-roll era the
function $f(\varphi (t)$ must be constant, or a slowly-varying
function of the cosmic time. In the following sections we shall
present two models for which this qualitative behavior occurs, but
in principle there exist a vast number of qualitative behaviors that
can be realized, depending the choice of $f(\varphi (t))$. As we
already stated earlier, the function $f(\varphi (t))$ must be a
monotonic function of the scalar field $\varphi$. Before we proceed
to the examples, it is important to derive certain relations that
will enable us to validate our findings. One important relation can
be found by combining Eqs.  (\ref{kleingordonsingle}),
(\ref{basiccondition}), (\ref{extra1}) and
(\ref{potentialhubblscalar}), and it is,
\begin{equation}\label{validitycheckrelation}
\dot{\varphi}=-\frac{V'(\varphi)}{H(\varphi)(f(\varphi)+3)}\, ,
\end{equation}
which in the slow-roll era is approximately equal to,
\begin{equation}\label{validitycheckrelation1}
\dot{\varphi}=-\frac{V'(\varphi)}{H(\varphi)(3)}\, ,
\end{equation}
Consider that we found a solution $H(\varphi)$ of the differential
equation (\ref{masterequation}), then if we find the approximate
expression for  $H(\varphi)$ during the slow-roll, the expressions
(\ref{validitycheckrelation}) and (\ref{validitycheckrelation1})
should coincide in the slow-roll limit. Another useful relation is
the $e$-foldings number $N$ as a function of $f(\varphi )$. The
$e$-foldings number is defined as follows,
\begin{equation}\label{efoldingsnumbern}
N=-\int_{t_k}^{t}H(t)\mathrm{d}t=-\int_{\varphi_k}^{\varphi}\frac{H}{\dot{\varphi}}\mathrm{d}\varphi
\, ,
\end{equation}
where $t_k$ is the time instance that the horizon crossing occurs,
and $\varphi_k$ is the value of the canonical scalar field at the
horizon crossing. By substituting Eq. (\ref{validitycheckrelation1})
in Eq. (\ref{efoldingsnumbern}), we obtain,
\begin{equation}\label{efoldings1}
N=\frac{1}{M_p^2}\int_{\varphi}^{\varphi_k}\frac{V(f(\varphi)+3)}{3V'(\varphi)}\mathrm{d}\varphi
\, .
\end{equation}
During the slow-roll era, the above reduces to the well-known
relation,
\begin{equation}\label{efoldings}
N\simeq \frac{1}{M_p^2}
\int_{\varphi}^{\varphi_{i}}\frac{V(\varphi)}{V'(\varphi)}\mathrm{d}\varphi
\, .
\end{equation}
In the next two sections we shall present two models which allow a
smooth transition between constant-roll and slow-roll eras, and we
shall discuss in some detail the theoretical implications  of each
model.

\section{Model I: From Slow-Rolling $\alpha$-attractors to Constant-Roll}

\begin{figure}[h]
\centering
\includegraphics[width=15pc]{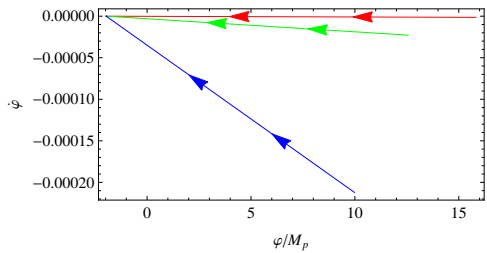}
\caption{Phase space structure of the solution
(\ref{generalsolution1}), for the initial conditions $\varphi
(0)=10^{20.2}$ (red line), $\varphi (0)=10^{20.1}$ (green line),
$\varphi (0)=10^{20}$ (blue line).}\label{plotsnumerics1}
\end{figure}

\begin{figure}[h]
\centering
\includegraphics[width=15pc]{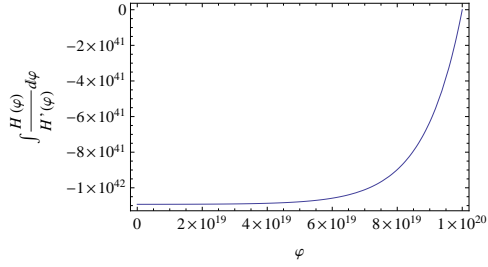}
\caption{Stability of linear perturbations for the solution
(\ref{generalsolution1}).}\label{stabilityperturbations}
\end{figure}

Let us assume that the function $f(\varphi (t))$ appearing in Eq.
(\ref{basiccondition}) is chosen as follows,
\begin{equation}\label{choice1}
f(\varphi (t))=e^{-\lambda \varphi (t)}\, ,
\end{equation}
which is monotonic with respect to the scalar field $\varphi (t)$.
We shall choose the parameter $\lambda$ for the needs of this
section to be $\lambda=\frac{\sqrt{\frac{2}{3 \alpha }}}{M_p}$,
where $\alpha <1$, but for the moment we shall leave this
unspecified, in order to provide a general result. For $\lambda$
chosen as quoted above, the function $f(\varphi (t))$ has a quite
interesting behavior, since it allows a slow-roll era when
$\frac{\varphi}{M_p}\gg 1$, but in the case that $f(\varphi) \sim
\mathcal{O}(1)$, the constant-roll inflationary era can be realized.
Indeed, for large field values, when $\varphi> M_p$, and for
$\alpha<1$, the function $f(\varphi (t))$ is approximately zero,
since the exponential decays quite fast. In order to have a clear
quantitative picture of the physics behind the choice
(\ref{choice1}), let us here quote some numerical examples. In Table
\ref{table1} we present the values of the function $f(\varphi)$, for
various values of the scalar field $\varphi$. As it can be seen, for
$\varphi >20 M_p$, we have $f(\varphi)\ll 1$, so effectively the
second slow-roll index satisfies $\eta \ll 1$, and therefore the
slow-roll era can be realized and it is actually realized as we
evince later on in this section. For $\varphi \simeq 0.1 \, M_p$,
the second slow-roll index is of the order $\eta \simeq
\mathcal{O}(1)$, so effectively the constant-roll era begins, and in
this case $\frac{\ddot{\varphi}}{2H\dot{\varphi}}\simeq 1$. This
scenario of constant-roll was described in \cite{Motohashi:2014ppa},
and it corresponds to $\alpha=-4$ by using the notation of
\cite{Motohashi:2014ppa}, however $\alpha$ in Ref.
\cite{Motohashi:2014ppa} appears in
$\frac{\ddot{\varphi}}{2H\dot{\varphi}}=-(3+\alpha)$. Interestingly
enough, the constant-roll scenario corresponding to
$f(\varphi)\simeq 1$ yields quite appealing results, since if the
power-spectrum is generated during this era, it can be compatible
with the observational data, as it was shown in Ref.
\cite{Motohashi:2017aob}. However, we shall not discuss the
qualitative features of this model, but we concentrate on the
transition between the slow-roll and constant-roll eras and in the
general solution of the cosmological equations for the function
$f(\varphi)$ being chosen as in Eq. (\ref{choice1}). In addition, we
shall investigate how the slow-roll era can rise through the
resulting attractor solution. The details for the constant-roll era,
that is, the scalar potential and stability of the attractor
solution can be found in Refs.
\cite{Motohashi:2014ppa,Motohashi:2017aob}.
\begin{table*}[h]
\small \caption{\label{table1} Values of the function $f(\varphi)$
for various values of the scalar field for the model
(\ref{choice1}).}
\begin{tabular}{@{}crrrrrrrrrrr@{}}
 Scalar Field Values $\quad$ & $\quad$Values of the function $f(\varphi)$
\\
$\varphi\simeq 100 M_p$ $\quad $ &  $\quad $ $f(\varphi)\simeq
2.26214\times 10^{-40}$
\\
$\varphi\simeq 50 M_p$ $\quad $ &  $\quad $ $f(\varphi)\simeq
1.50404\times 10^{-20}$
\\ $\varphi\simeq 20 M_p$ $\quad $ &  $\quad $ $f(\varphi)\simeq
1.17735\times 10^{-8}$
\\ $\varphi\simeq  M_p$ $\quad $ &  $\quad $ $f(\varphi)\simeq
0.40137$
\\ $\varphi\simeq 0.1 M_p$ $\quad $ &  $\quad $ $f(\varphi)\simeq
0.912756$
\\ $\varphi\simeq 0.01 M_p$ $\quad $ &  $\quad $ $f(\varphi)\simeq
0.990913$
\\ $\varphi\simeq 0.0001 M_p$ $\quad $ &  $\quad $ $f(\varphi)\simeq
0.999909$\\ $\varphi\simeq 10^{-7} M_p$ $\quad $ &  $\quad $
$f(\varphi)\simeq 1$\\
\end{tabular}
\end{table*}
Let us focus on the general case with $f(\varphi)$ being chosen as
in Eq. (\ref{choice1}), so by solving the differential equation
(\ref{masterequation}), we obtain the following solution $H(\varphi
)$,
\begin{equation}\label{generalsolution1}
H(\varphi)=C_1\, J_0\left(\frac{\left(2 \sqrt{\beta }\right)
e^{-\frac{1}{2} (\lambda  \varphi )}}{\lambda }\right)\, ,
\end{equation}
where $J_0(x)$ is the Bessel function of the first kind, $C_1$ is an
integration constant and $\beta=\frac{1}{2M_p^2}$. The corresponding
scalar potential can easily be found by substituting $H(\varphi )$
from Eq. (\ref{generalsolution1}) in Eq.
(\ref{potentialhubblscalar}), so the result is,
\begin{equation}\label{potentialcase1}
V(\varphi)=3 C_1^2 M_p^2J_0\left(\frac{\left(2 \sqrt{\beta }\right)
e^{-\frac{1}{2} (\lambda  \varphi )}}{\lambda }\right)^2-2 \beta
C_1^2 M_p^4 e^{-\lambda  \varphi }J_1\left(\frac{\left(2 \sqrt{\beta
}\right) e^{-\frac{1}{2} (\lambda  \varphi )}}{\lambda }\right)^2\,
.
\end{equation}
At this point we shall make a numerical investigation in order to
see whether the solution (\ref{generalsolution1}) is an attractor of
the cosmological equations, independently of the initial conditions.
By using the following values of the free parameters (recall that
$\beta=\frac{1}{2M_p^2}$), for reasons that will become clear later
on,
\begin{equation}\label{choicesofparameters}
\lambda=\frac{\sqrt{\frac{2}{3 \alpha
}}}{M_p},\,\,\,\alpha=0.9,\,\,\,C_1=1\, ,
\end{equation}
in Fig. \ref{plotsnumerics1} we plot the phase space behavior of the
solution (\ref{generalsolution1}), for the initial conditions
$\varphi (0)=10^{20.2}$ (red line), $\varphi (0)=10^{20.1}$ (green
line), $\varphi (0)=10^{20}$ (blue line).
At it can be seen, all three solutions converge to a unique
attractor, so the solution (\ref{generalsolution1}) is the unique
attractor of the cosmological dynamical system. More importantly,
the attractor occurs as the scalar field values decrease, and at the
same time the velocity of the scalar decreases. Eventually the
scalar field settles at some value, when the velocity becomes zero.
This can also be verified analytically for the limiting slow-roll
case by using Eq. (\ref{perturbationsolution1}), but we do this
later on. However, we can make a numerical analysis of the solution
(\ref{perturbationsolution1}), by using $H_0(\varphi)$ as in Eq.
(\ref{generalsolution1}). In Fig. \ref{stabilityperturbations} we
plot the behavior of the exponential in Eq.
(\ref{perturbationsolution1}), which determines the behavior of the
perturbations, for $\varphi$ being chosen in the interval
$\varphi=(10^{10},10^{21})$. As it can be seen, for the values of
interest, the linear perturbations decay quite rapidly, so this also
verifies that the solution (\ref{generalsolution1}) is actually the
solely attractor of the cosmological system.

Let us now focus on the slow-roll era, which corresponds to
$\varphi>20\,M_p$, in which case the arguments of the Bessel
functions appearing in Eq. (\ref{potentialcase1}) are very small. So
the potential (\ref{potentialcase1}) can be approximated as follows,
\begin{equation}\label{approxpot}
V(\varphi)\simeq -\frac{2 \beta ^2C_1^2 M_p^4 e^{-2 \lambda \varphi
}}{\lambda ^2}-\frac{6 \beta  C_1^2 M_p^2 e^{-\lambda \varphi
}}{\lambda ^2}+3 C_1^2 M_p^2\, .
\end{equation}
The above potential has very interesting structure since it is quite
similar to the Starobinsky model and the $\alpha$-attractor models
at leading order. Actually, as we shall evince shortly, the
resulting observational indices, that is, the spectral index of the
power spectrum of primordial curvature perturbations, and the
scalar-to-tensor ratio corresponding to the potential
(\ref{approxpot}) are identical to the ones corresponding to the
$\alpha$-attractor models. Before proceeding to the calculation of
the spectral index and of the scalar-to-tensor ratio, let us
validate that the expressions for $\dot{\varphi}$ appearing in Eqs.
(\ref{validitycheckrelation}) and (\ref{validitycheckrelation1}) are
identical in the slow-roll limit. By substituting $H(\varphi)$ from
Eq. (\ref{generalsolution1}) in Eq. (\ref{validitycheckrelation1}),
we obtain,
\begin{equation}\label{dotphislowrolldirect}
\dot{\varphi}= -2 \sqrt{\beta } C_1 M_p^2 e^{-\frac{1}{2} \lambda
\varphi } J_1\left(\frac{2 e^{-\frac{1}{2} \lambda \varphi }
\sqrt{\beta }}{\lambda }\right)\, ,
\end{equation}
which in the slow-roll limit reads,
\begin{equation}\label{dotphislowrolldirect1}
\dot{\varphi}\simeq -\frac{2 \beta  C_1 M_p^2 e^{-\lambda \phi
}}{\lambda }\, .
\end{equation}
Accordingly, by substituting (\ref{generalsolution1}) in Eq.
(\ref{validitycheckrelation}), we get,
\begin{equation}\label{dotphislowrollindicrerct}
\dot{\varphi}=\frac{2 \beta  M_p^2 e^{-\lambda  \varphi } \,
_0\tilde{F}_1\left(;2;-\frac{e^{-\lambda  \varphi } \beta }{\lambda
^2}\right) \left(3 e^{\lambda  \varphi }+2 \beta
M_p^2\right)}{\lambda  \left(\beta -3 e^{\lambda  \varphi
}\right)}\, ,
\end{equation}
where $_0\tilde{F}_1(;z,x)$ is the regularized hypergeometric
function. By taking the slow-roll limit, the expression in Eq.
(\ref{dotphislowrollindicrerct}) becomes at leading order,
\begin{equation}\label{dotphislowrollindicrerct}
\dot{\varphi}\simeq -\frac{2 \beta  C_1 M_p^2 e^{-\lambda \phi
}}{\lambda }\, ,
\end{equation}
which is identical with the expression appearing in Eq.
(\ref{dotphislowrolldirect1}). This result validates that the
solution (\ref{generalsolution1}) yields the correct slow-roll limit
for the cosmological system. Also it is worth finding the analytic
behavior of the perturbations $\delta H$ appearing in Eq.
(\ref{perturbationsolution1}) during the slow-roll era. The Hubble
rate (\ref{generalsolution1}) during the slow-roll era can be
approximated as,
\begin{equation}\label{hubblesloworollapprox}
H(\varphi)\simeq 1-\frac{\beta  e^{-\lambda \varphi }}{\lambda ^2}\,
,
\end{equation}
so by substituting in Eq. (\ref{perturbationsolution1}), and by
integrating with $\varphi$ in the interval $(\varphi_k,\varphi_f)$,
the resulting expression for the evolution of linear perturbations
read,
\begin{equation}\label{perturbationsolution1}
\delta H(\varphi )\simeq \delta
H(\varphi_0)e^{\frac{3}{2M_p^2}\left( \frac{e^{\lambda
\varphi_f}}{\beta }-\frac{e^{\lambda  \varphi_k}}{\beta }\right)}\,
,
\end{equation}
where $\varphi_f$ is the value of the scalar field at the time
instance that the slow-roll era ends. Since $\varphi_f\ll
\varphi_k$, the exponent of the exponential is negative, and
therefore the perturbations $\delta H$ decay in a double exponential
way, so quite fast. This result verifies analytically the numerical
investigation of Fig. \ref{stabilityperturbations}, and therefore
the solution $H(\varphi)$ is the attractor of the cosmological
equations, even during the slow-roll era.

Now let us proceed to the calculation of the observational indices
for the scalar potential (\ref{approxpot}). The study of the first
slow-roll index will reveal when the slow-roll era actually ends.
The Table \ref{table1} gives us a first idea on when inflation ends,
but only the study of the first slow-roll index will reveal the
actual ending of the slow-roll era. However, as we see, the
situation is perplexed since the function $f(\varphi)$ is already
non-zero at $\varphi_f$, hence the approximation starts to collapse.
It is therefore possible that the constant-roll takes over before
$\varphi_f$, so some $e$-folds are done within this constant-roll
era. The complete analysis however requires a more detailed
numerical approach and it exceeds the purposes of this article so we
defer this task to a future work.

During the slow-roll era, the slow-roll indices can be calculated by
using the scalar potential, by using the following slow-roll
formulas,
\begin{equation}\label{slowrollscalar}
\epsilon (\varphi)=\frac{M_p^2}{2}\left(
\frac{V'(\varphi)}{V(\varphi)}\right)^2,\,\,\,\eta
(\varphi)=M_p^2\frac{V''(\varphi)}{V(\varphi)}\, ,
\end{equation}
so by using the scalar potential (\ref{potentialcase1}), the first
slow-roll index during the slow-roll era,evaluated at the horizon
crossing, reads,
\begin{equation}\label{firstslowroll}
\epsilon \simeq \frac{2 \beta ^2 \lambda ^2 M_p^2}{\left(\lambda ^2
e^{\lambda  \varphi_k }-2 \beta \right)^2}\, ,
\end{equation}
It is vital for the following to calculate the exact value of the
scalar field for which $\epsilon \sim \mathcal{O}(1)$, so by solving
$\epsilon \sim 1$, we get,
\begin{equation}\label{epsilonoforderone}
\varphi_f \simeq \frac{\ln \left(\frac{2 \beta +\sqrt{2} \beta
\lambda M_p}{\lambda ^2}\right)}{\lambda }\, .
\end{equation}
In order to have a quantitative idea of the magnitude $\varphi_f$,
we choose the parameters as in Eq. (\ref{choicesofparameters})
(recall $\beta=\frac{1}{2M_p^2}$), so $\varphi_f$ becomes
$\varphi_f\simeq 7.45\times 10^{18}$. By looking at Table
\ref{table1}, we can see that for this value, the function
$f(\varphi)$ is $f(\varphi)\simeq 0.541$, which is very close to
$\mathcal{O}(1)$, but still not one. Hence, it is debatable whether
the constant-roll era starts at $\varphi_f$. However, we assume here
that the slow-roll era stops at $\varphi_f$, so after that point the
constant-roll era takes over and some $e$-foldings are done during
this era. The exit from inflation can be an issue in these
constant-roll theories however. In principle the exit from inflation
is controlled by the second slow-roll index $\eta$, so when this
ceases to be small, exit comes, however a numerical analysis must be
performed, and this will be done elsewhere.

During the slow-roll era, the $e$-foldings number $N$ can be
expressed in terms of the potential $V(\varphi)$,
\begin{equation}\label{efoldings}
N\simeq \frac{1}{M_p^2}
\int_{\varphi_f}^{\varphi_{i}}\frac{V(\varphi)}{V'(\varphi)}\mathrm{d}\varphi
\, ,
\end{equation}
where $\varphi_i$ is an initial value of the scalar field, which we
assume it to be the value at the horizon crossing $\varphi_k$. By
substituting the potential (\ref{potentialcase1}) we get at leading
order,
\begin{equation}\label{leadingordern}
N\simeq \frac{e^{\lambda  \varphi_k}-e^{\lambda  \varphi_f}}{2 \beta
M_p^2}\, .
\end{equation}
By substituting $\varphi_f$ from Eq. (\ref{epsilonoforderone}) in
Eq. (\ref{leadingordern}), we obtain $\varphi_k$ in terms of $N$,
\begin{equation}\label{phikleadingorder}
\varphi_k\simeq \frac{\log \left(\frac{\beta  \left(2 \lambda ^2
M_p^2 N-\sqrt{2} \lambda  M_p+2\right)}{\lambda ^2}\right)}{\lambda
}\, .
\end{equation}
Hence by substituting (\ref{phikleadingorder}) in Eq.
(\ref{firstslowroll}), the first slow-roll index becomes,
\begin{equation}\label{firstslowrollfinall}
\epsilon \simeq \frac{2}{\left(\sqrt{2}-2 \lambda M_p N\right)^2}\,
.
\end{equation}
Accordingly, the second slow-roll index at leading order, evaluated
at the horizon crossing, is approximately equal to,
\begin{equation}\label{etaathorizon}
\eta \simeq -2 \beta  M_p^2 e^{-\lambda  \varphi_k }\, ,
\end{equation}
so by substituting $\varphi_k$ from Eq. (\ref{phikleadingorder}), we
get,
\begin{equation}\label{etaleadingorderprefinal}
\eta \simeq -\frac{2 \lambda ^2 M_p^2}{2 \lambda ^2 M_p^2 N-\sqrt{2}
\lambda  M_p+2}\, .
\end{equation}
The spectral index of primordial curvature perturbations and the
scalar-to-tensor ratio in terms of the slow-roll indices, during the
slow-roll era, read,
\begin{equation}\label{spectscalindex}
n_s\simeq 1-6\epsilon+2\eta,\,\,\, r\simeq 16 \epsilon\, ,
\end{equation}
so by substituting Eqs. (\ref{firstslowrollfinall}) and
(\ref{etaleadingorderprefinal}) in Eq. (\ref{spectscalindex}) and by
using the $\lambda$ as in Eq. (\ref{choicesofparameters}), we get at
leading order for large $N$,
\begin{equation}\label{osrik}
n_s\simeq 1-\frac{2}{N}-\frac{9}{2 N^2},\,\,\,r\simeq \frac{12
\alpha }{N^2}\, .
\end{equation}
These are identical to the $\alpha$-attractors observational
indices, so we are confronted with this interesting scenario of
having a slow-roll era which lasts approximately $50-60$ $e$-folds,
which is followed immediately by a constant-roll era. Then it is
possible that the constant-roll era takes over and controls the
dynamics, as this was described in Refs.
\cite{Martin:2012pe,Motohashi:2014ppa,Motohashi:2017aob}. Then one
could speculate that non-Gaussianities can be generated, since their
three-point functions depend on the slow-roll indices. Actually, by
using the results of \cite{Martin:2012pe}, a rough estimate of the
parameter $f_{NL}$ for the constant-roll scenario we describe in
this paper is $f_{NL}\simeq 7.5$, which can be verified or excluded
by future observations. Actually this value is within the current
bound $-10<f_{NL}<110$, if $f_{NL}$ is assumed to be momentum
independent, which corresponds to local non-Gaussianities. However,
one should be cautious with regards to the three-point functions,
since the three-point functions are calculated in a conformal time
interval, for which the initial time corresponds to an era that the
corresponding $f_k$ modes are well inside the Hubble radius during
the inflationary era. In this paper we approximated the potential
during the slow-roll era, so we cannot be sure that if we use these
semi-analytic approximations, we shall get accurate results. An
accurate approach requires a full analytic treatment, or a detailed
numerical treatment, which however exceeds the purposes of this
work.

\begin{figure}[h]
\centering
\includegraphics[width=15pc]{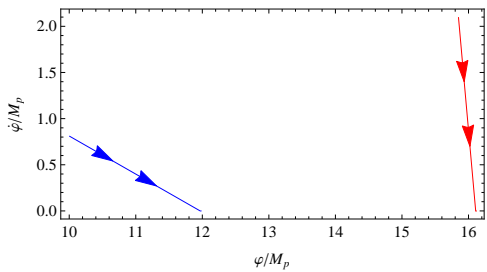}
\caption{Phase space structure of the solution
(\ref{hfmodelvariant1}). The red curve corresponds to the initial
condition $\varphi (0)=10^{20.2}$ and the blue curve to $\varphi
(0)=10^{20}$.}\label{newnumerics1}
\end{figure}

Before closing we need to note that it is possible to construct a
variant model of the one appearing in Eq. (\ref{choice1}), for which
the constant-roll era occurs before the slow-roll era. For example,
if we choose the function $f(\varphi )$ as follows,
\begin{equation}\label{asx1}
f(\varphi )=e^{-\lambda (\varphi-x_0)}\, ,
\end{equation}
with $x_0\simeq M_p$. Then, the resulting solution of the
differential equation (\ref{masterequation}) is,
\begin{equation}\label{hfmodelvariant1}
H(\varphi)=C_1J_0\left(\frac{2 \sqrt{\beta } \sqrt{e^{-\lambda
(x_0-\varphi)}}}{\lambda }\right)\, .
\end{equation}
The corresponding scalar potential is,
\begin{equation}\label{asxetopotential1}
V(\varphi )=3 M_p^2 J_0\left(\frac{2 e^{\frac{1}{2} \lambda (\varphi
-x_0)} \sqrt{\beta }}{\lambda }\right){}^2-2 \beta M_p^4 e^{\lambda
(\varphi -x_0)} J_1\left(\frac{2 e^{\frac{1}{2} \lambda (\varphi
-x_0)} \sqrt{\beta }}{\lambda }\right){}^2\, .
\end{equation}
We performed a detailed numerical analysis for the solution
(\ref{hfmodelvariant1}), and the results appear in Fig.
(\ref{newnumerics1}) in which we plot the phase space evolution
corresponding to the solution (\ref{hfmodelvariant1}). We use the
values $x_0\simeq M_p$, and the rest of the parameters as in Eq.
(\ref{choicesofparameters}), with $\beta=\frac{1}{2M_p^2}$. The red
curve corresponds to the initial condition $\varphi (0)=10^{20.2}$
and the blue curve to $\varphi (0)=10^{20}$.
As it can be seen in Fig. (\ref{newnumerics1}), the solution is very
sensitive to initial conditions, and therefore the solution
(\ref{hfmodelvariant1}) is not an attractor of the cosmological
dynamics. Hence we refer from going into details for this case.
However, the feature we just described, that is, the non-attractor
feature that is met when the constant-roll occurs before inflation,
is very model dependent.

%%%%%%%%%%%%%%%%%%%%%%%%%%%%%%%%%%%%%%%%%%%%%%%%%%%%%%%%%%%%%%%%%%%%%%%%%%%%%%%%%%%%%%%%%%%%%%%%%%%%%%%%%

\section{Constant-Roll Before the Slow-Roll Era: A Toy Model}

\begin{figure}[h]
\centering
\includegraphics[width=15pc]{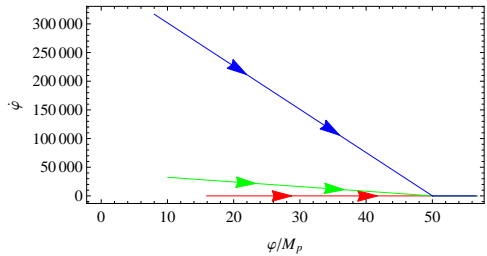}
\caption{Phase space structure of the solution
(\ref{hubbleratechoice2sol}), for the initial conditions $\varphi
(0)=10^{20.2}$ (red line), $\varphi (0)=10^{20}$ (green line),
$\varphi (0)=10^{19.9}$ (blue line).}\label{akyroplot}
\end{figure}

\begin{figure}[h]
\centering
\includegraphics[width=15pc]{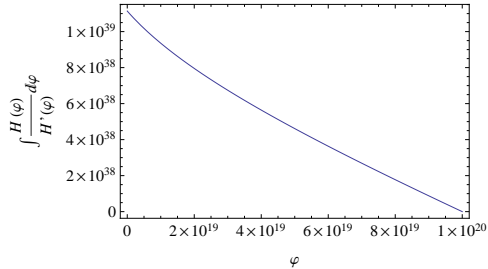}
\caption{Evolution of linear perturbations of the solution
(\ref{hubbleratechoice2sol}).}\label{numericssecondmodel}
\end{figure}

Let us now assume that the function $f(\varphi (t))$  in Eq.
(\ref{basiccondition}) has the following form,
\begin{equation}\label{choice2}
f(\varphi )=3\,e^{-\frac{\lambda  \varphi}{M_p}}-3\, ,
\end{equation}
where $\lambda$ is a positive number which can be arbitrary. Then,
the function $f(\varphi )$ for $\varphi > M_p$ is $f(\varphi)\simeq
-3$ , and for $\varphi\ll M_p$, it approaches zero. Thus in this
scenario, the constant-roll era occurs before the slow-roll era, and
in this section we examine in brief the qualitative features of this
scenario. In Table \ref{table2} we present some characteristic
values of the function $f(\varphi)$ for various values of the scalar
field $\varphi$.
\begin{table*}[h]
\small \caption{\label{table2} Values of the function $f(\varphi)$
for various values of the scalar field for the model
(\ref{choice2}).}
\begin{tabular}{@{}crrrrrrrrrrr@{}}

 Scalar Field Values $\quad$ & $\quad$Values of the function $f(\varphi)$
\\
$\varphi\simeq 100 M_p$ $\quad $ &  $\quad $ $f(\varphi)\simeq -3$
\\
$\varphi\simeq 10 M_p$ $\quad $ &  $\quad $ $f(\varphi)\simeq -3$
\\
$\varphi\simeq 5 M_p$ $\quad $ &  $\quad $ $f(\varphi)\simeq
-2.99986$
\\ $\varphi\simeq  M_p$ $\quad $ &  $\quad $ $f(\varphi)\simeq
-2.59399$
\\ $\varphi\simeq 0.01 M_p$ $\quad $ &  $\quad $ $f(\varphi)\simeq
-0.059404$
\\ $\varphi\simeq 10^{-6} M_p$ $\quad $ &  $\quad $ $f(\varphi)\simeq
-5.99999\times 10^{-6}$
\\ $\varphi\simeq 10^{-10} M_p$ $\quad $ &  $\quad $ $f(\varphi)\simeq
-6\times 10^{-10}$
\\

\end{tabular}
\end{table*}
As it can be seen, for $\varphi >10 M_p$, the constant-roll era
occurs, and for $\varphi\preceq  10^{-6} M_p$, the slow-roll era
takes over. Hence, for large field values, the constant-roll era
occurs, and the cosmological system continuously deforms to the
slow-roll era, which takes place for small values of the scalar
field. Let us note that the case constant-roll case described by Eq.
(\ref{choice2}) describes the $\alpha=0$ case of Ref.
\cite{Motohashi:2014ppa}, or the $n=-3$ case of Ref.
\cite{Martin:2012pe}, which is called ultra-slow-roll in that work.
Naively, one could claim that the evolution is controlled by the
dynamics of constant-roll up to the point that the field values
become of the order $\varphi \simeq 10^{-6} M_p$, however this is
not so as we now evince. This could be true if one sees the
cosmological equations in their constant-roll limit or slow-roll
limit, but if we use the dynamical approach quantified by the
function $f(\varphi (t))$, the resulting picture is different. This
is due to the fact that the solution of Eq. (\ref{masterequation})
must be the attractor of the cosmological system, and also the
correct attractor, that is, the phase space curves must tend to a
phase space point for which the velocity of the field is zero, as
the scalar field values decrease. Eventually the ideal situation is
that the velocity becomes zero at the attractor point, which shows
that the field settles at a point as it rolls down to its potential.
This is not the case for the model (\ref{choice2}) as we now evince.
The solution of the differential equation (\ref{masterequation}) for
the model (\ref{choice2}) is,
\begin{equation}\label{hubbleratechoice2sol}
H(\varphi)=C_1 \Gamma \left(1+\frac{\sqrt{6}}{\lambda }\right)
J_{\frac{\sqrt{6}}{\lambda }}\left(\frac{\sqrt{6} e^{-\frac{\varphi
\lambda }{2 M_p}}}{\lambda }\right)\, .
\end{equation}
It can be shown that in the slow-roll limit, the leading order
potential is,
\begin{equation}\label{leadingorderchoice2potential}
V(\varphi)\simeq \delta e^{-\frac{\sqrt{6} \varphi }{M_p}}\, ,
\end{equation}
where $\delta$ is a constant parameter which is positive but we do
not quote it's analytic form for brevity. The full form of the
potential corresponding to the solution
(\ref{leadingorderchoice2potential}) is found by substituting
(\ref{leadingorderchoice2potential}) in Eq.
(\ref{potentialhubblscalar}), and we get,
\begin{align}\label{fullpotentialfordifficultcase}
& V(\varphi)=3 C_1^2 M_p^2 \Gamma \left(1+\frac{\sqrt{6}}{\lambda
}\right)^2 J_{\frac{\sqrt{6}}{\lambda }}\left(\frac{\sqrt{6}
e^{-\frac{\lambda \varphi }{2 M_p}}}{\lambda }\right){}^2\\
\notag &-\frac{3}{4} C_1^2 M_p^2 \Gamma
\left(1+\frac{\sqrt{6}}{\lambda }\right)^2 e^{-\frac{\lambda \varphi
}{M_p}} \left(J_{\frac{\sqrt{6}}{\lambda }-1}\left(\frac{\sqrt{6}
e^{-\frac{\lambda  \varphi }{2 M_p}}}{\lambda
}\right)-J_{1+\frac{\sqrt{6}}{\lambda }}\left(\frac{\sqrt{6}
e^{-\frac{\lambda  \varphi }{2 M_p}}}{\lambda }\right)\right){}^2\,
.
\end{align}
Let us make a numerical investigation in order to validate whether
the solution (\ref{hubbleratechoice2sol}) is an attractor of the
cosmological equations in this case. We use the following values of
the free parameters, without loss of generality, $\lambda=0.2$,
$C_1=1$, and in Fig. \ref{akyroplot} we plot the phase space
behavior of the solution (\ref{hubbleratechoice2sol}), for the
initial conditions $\varphi (0)=10^{20.2}$ (red line), $\varphi
(0)=10^{20}$ (green line), $\varphi (0)=10^{19.9}$ (blue line).
As it can be seen in Fig. \ref{akyroplot}, there is an attractor
behavior, since all the solutions converge to the same point,
however, the attractor occurs in the opposite direction from the
desirable one. Indeed, as the field value decreases, the velocity of
the scalar field grows, and this is not a desirable feature. So it
seems that the solution (\ref{hubbleratechoice2sol}) is not an
attractor of the cosmological system. In order to further support
this argument, we shall investigate the evolution of linear
perturbations, as these are given by Eq.
(\ref{perturbationsolution1}), for the solution
(\ref{hubbleratechoice2sol}). The results of our analysis appear in
Fig. \ref{numericssecondmodel}, and as it can be seen, the linear
perturbations from the solution (\ref{hubbleratechoice2sol}) grow
very rapidly. Hence, this further supports our argument that the
solution (\ref{hubbleratechoice2sol}) is not an attractor of the
cosmological dynamics for the model (\ref{choice2}). We need to note
here that this non-attractor behavior related to a constant second
slow-roll index $\eta$ is a quite well known phenomenon in the
literature. Particularly, the ultra-slow-roll solution of Ref.
\cite{Kinney:2005vj}, is an inherently non-attractor solution.
Another related scenario describing a constant-roll to a slow-roll
transition was presented in \cite{Tzirakis:2007bf}, in which case,
the initial constant-roll phase is a dynamically transient phase. In
fact, this scenario has the interesting feature of the constant-roll
era occurring before the slow-roll era. This behavior leaves space
for a possible graceful exit from inflation, when the slow-roll era
ends. We hope to also address this behavior in a future work.

Now the question is if the above behavior occurs in all the cases
that the slow-roll era occurs after the constant-roll era, at least
in the context of the dynamical mechanism we describe in this paper.
By observing the two models we presented so far, namely
(\ref{choice1}) and (\ref{choice2}), the answer is yes. However,
this is not true in the general case, unless it is proved by using
generic arguments and in a concrete way. For the moment let us note
that this feature is model dependent, always in the context of the
dynamical approach we used in this paper. For example, the model
(\ref{newchoice2}) below, leads to attractor behavior, and the
constant-roll era occurs before the slow-roll era.

Before closing we would like to note that the mechanism we propose
allows for different transition scenarios, and here we give a couple
of interesting cases, which will be addressed elsewhere. An
interesting and dynamically stable choice for the function
$f(\varphi )$ is the following,
\begin{equation}\label{newchoice1}
f(\varphi )=-\frac{1}{\lambda \frac{\varphi}{M_p}+\beta}\, ,
\end{equation}
where $\lambda$ is a positive dimensionless constant and the same
applies for  $\beta$. In the model described by (\ref{newchoice1}),
for large field values $\varphi>M_p$, the function $f(\varphi )$
becomes $f(\varphi )\sim 0$ and for small field values, that is
$\varphi\ll M_p$, we have $f(\varphi )\simeq -\frac{1}{\beta}$.
Therefore, the slow-roll era can be realized for large field values,
and the constant-roll for small field values. As it can be shown,
the model (\ref{newchoice1}) leads to attractor behavior, and the
resulting potential during the slow-roll era, is a power-law
potential. A variant form of the model (\ref{newchoice1}) is the
following,
\begin{equation}\label{newchoice2}
f(\varphi )=-\frac{1}{\lambda (\varphi-\varphi_0)+\beta}\, ,
\end{equation}
in which case the attractor behavior persists, and also
constant-roll occurs before the slow-roll era.

Also, with the dynamical mechanism we proposed, it is possible to
have transitions between constant-roll eras. For example if we make
the following choice,
\begin{equation}\label{fconstantrolltransitions}
f(\varphi)=-\frac{n}{\gamma\, e^{-\lambda \varphi}+\delta}\, ,
\end{equation}
where $\lambda$ is a positive constant with dimensions $M_p^{-1}$,
and $\delta>0$, then for large field values, a constant roll era is
realized and for small field values, the result is a different
constant-roll era. The models (\ref{newchoice2}) and
(\ref{fconstantrolltransitions}), are quite stable and we aim to
study these in some detail elsewhere.

In principle there are more choices for the function $f(\varphi)$,
with the most interesting and easy to handle being functions that
contain exponentials. Another interesting choice is to use
trigonometric functions, so a kind of oscillating behavior would
occur, but this study exceeds the purpose of this introductory work.

\section{Concluding Remarks}

In this work we studied the implications on the dynamics of
inflation of a dynamically varying second slow-roll index $\eta$.
Particularly we assumed that the second slow-roll index is equal to
a function of the canonical scalar field $f(\varphi )$, and we
focused on models for which the function, in certain limiting values
of the scalar field, becomes either very small or a constant number.
The cases that the function is very small correspond to a slow-roll
era, and the cases where the function is a constant corresponds to a
constant-roll era. This mechanism we studied shows how it is
possible to have a transition from a slow-roll to a constant-roll
era. As particular examples we studied models which contain
exponentials in the function $f(\varphi )$, with the most
interesting case leading to a slow-roll era which is characterized
by an $\alpha$-attractor evolution, which is followed by a
constant-roll evolution. The constant-roll era starts for a value of
the scalar field very close to the value for which the first
slow-roll index becomes of the order one. We then speculated that it
is possible for the cosmological system to follow the constant-roll
attractor for some $e$-foldings, in which case certain amount of
non-Gaussianities could be generated. The stability of the attractor
solution was also checked and we found that the solution is stable
and an attractor of the cosmological system. Also we examined the
possibility that the constant-roll occurs before the slow-roll era,
but the resulting solutions were unstable. As we also commented in
the main text, this issue is probably highly model dependent and
need to be further studied.

Having a smooth transition from a slow-roll era to a constant-roll
era has some attributes but also some drawbacks. For example, the
primordial perturbations can be generated during the slow-roll era,
and also these do not evolve after the horizon crossing. Then, after
the Universe enters the constant-roll era, the non-Gaussianities may
be generated. However with this work we did not address in detail
this issue, and this has to be done by using the full solution
derived by our approach. The analytic treatment however can be
tedious, so in a future work we hope to address this at least
numerically to some extent. Specifically, if the theory is treated
as it behaves in its various limits, it is easy to address all the
issues, but if we use the exact analytic solution, the calculations
are much more difficult and the various eras cannot be easily
distinguished. Therefore only a focused numerical analysis can
reveal the true dynamics of our approach. One issue that also needs
to be consistently addressed is the graceful exit from inflation.
Particularly, usually this exit is identified with the time instance
that the first slow-roll index becomes of the order one. However,
this indicates that the slow-roll era ends, so if a constant-roll
era follows the slow-roll era, the question is then, when does
actually inflation ends. This issue is of profound importance since
during the constant-roll era, the perturbations generated evolve
dynamically, and also the reheating process must start eventually.
We hope to address some of these issues elsewhere.

\section*{Acknowledgments}

This work is supported by MINECO (Spain), project
 FIS2013-44881, FIS2016-76363-P and by CSIC I-LINK1019 Project (S.D.O) and by Min. of Education and Science of Russia (S.D.O
and V.K.O).

\end{document}